\documentclass[11pt,preprint,aps,prc,showpacs,showkeys]{revtex4}
\usepackage[dvips]{graphicx}
\newcommand{\be}{\begin{eqnarray}}
\newcommand{\ee}{\end{eqnarray}}

\newcommand{\nd}{\noindent}
\begin{document}
\title{$J/\psi$ suppression at forward rapidity as a potential probe for QGP formation in colour screening scenario}
\vskip 0.2in
{\author{M. Mishra}
\email{madhukar.12@gmail.com} 
\author{C. P. Singh}
\email{cpsingh_bhu@yahoo.co.in}
\author{V. J. Menon}
\affiliation{Department of Physics, Banaras Hindu University, Varanasi 221005, India}
\vskip 0.1in
\begin{abstract}
   In order to study the properties of $J/\psi$ (1S) in the deconfining medium, we extend our previous formalism~[Phys. Lett. B {\bf 656}, 45 (2007)] on $J/\psi$ suppression at mid-rapidity using the colour screening framework. Our formalism is more general as the complete rapidity, transverse momentum and centrality dependence including $J/\psi$ suppression at forward as well as mid-rapidity can be computed directly from it. Careful attention is paid to the role of the medium's proper time in determining the locus of the screening region where $J/\psi$ gets suppressed. Other important ingredients in the calculation are bag model equation of state for QGP, the longitudinal expansion of the QGP fluid obeying Bjorken's boost invariant scaling law and non-sequential/sequential melting of $\chi_c$ (1P) as well as $\psi^{'}$ (2S) higher resonances. Upon comparison with the recent data of PHENIX collaboration on $J/\psi$ suppression at forward and mid-rapidity regions, we find that our model shows a reasonable agreement with the data without incorporating any sequential decay mechanism of higher charmonia states.  Furthermore, we observe a larger suppression at forward rapidity in our model which is again well supported by the PHENIX data and also gives a hint that a scenario based on directly produced $J/\psi$'s is preferable.      
\end{abstract}  
\vskip 0.2in
\pacs{25.75.-q; 25.75.Nq; 12.38.Mh; 25.75.Gz}
\keywords{$J/\psi$ suppression; survival probability; sequential melting; relativistic heavy-ion collisions; colour Debye screening; general rapidity.}

\maketitle

\section{Introduction}
    The current experimental programmes at RHIC are focused towards the investigations of the properties of hadrons or hadronic resonances above the deconfinement transition. In particular, suppression of $J/\psi$ (or charmonium state) has been suggested as a potential probe of the deconfined matter produced in the ultra-relativistic heavy-ion collisions. Lattice quantum chromodynamics (QCD) calculations reveal that the critical temperature for confined normal nuclear matter (HG) to deconfined matter (QGP) phase transition is $T_c\sim 0.160-0.190$ GeV if baryon chemical potential $\mu_B=0$ \cite{hats} although there are controversies whether such a transition is a first order or only a crossover~\cite{fod,kar}. Matsui and Satz~\cite{mats} first predicted that the binding potential of the $c\bar c$ pair into $J/\psi$ mesons is screened in the presence of a QGP medium and $J/\psi$ states dissociate at temperatures for which the colour (Debye) screening radius of the medium falls below their corresponding $c\bar c$ binding radius. A crucial assumption in favour of QGP is that charmonia once dissociated in the medium cannot recreated at the hadronization stage. Thermodynamics of equilibrated QGP does not allow $c\bar c$ to be created in abundance and hence there is a suppression in $J/\psi$ yields. The strength of the suppression depends on the binding energies of the quarkonia and the temperature of the medium~\cite{mats}. Old lattice QCD simulations~\cite{datt,satz} suggested that the $J/\psi$ (1S) could survive in QGP up to dissociation temperature $T_D\sim 2.1 T_c$ and the higher resonances, such as $\chi_c$ (1P) and $\psi^{'}$ (2S) can still melt near $T_c$. However new QCD analysis~\cite{amocsy} shows that the dissociation temperatures of all the relevant quarkonia viz $J/\psi,\,\chi_c,$ and $\psi^{'}$ are close to $T_c$. The $J/\psi$ and their higher resonances are usually created at the initial (pre-thermal) stage of heavy-ion collisions because of their large masses. Their small widths also make them almost insensitive to the final state interactions. Therefore, $J/\psi$ quarkonia can probe the evolution of the deconfined matter  beginning from the early stage of collisions~\cite{tann}.

   There are now high-statistics data available on the $J/\psi$ suppression obtained by NA50 experiments~\cite{abreu,aless,nac} at SPS and by PHENIX experiments at RHIC~\cite{pher}. One of the surprising result is the  observation of a similar suppression pattern at both these energies involving a large difference in the energy densities produced between SPS and RHIC. The other important feature of the PHENIX $J/\psi$ suppression data from Au$+$Au collisions at the center-of-mass energy $\sqrt{s_{NN}}=200$ GeV at RHIC is that $J/\psi$ yield in the central Au$+$Au collisions is suppressed by a factor of nearly 4 at mid-rapidity and 5 at forward rapidity as compared to that  observed in p$+$p collisions scaled by the average number of binary collisions. The absolute suppression of $J/\psi$ in heavy-ion collisions can arise due to initial state effects (e.g., cold nuclear matter effects) and/or final state effects ( e.g., colour screening arising due to possible formation of hot QGP). Two types of models have been proposed in order to explain the similar suppression pattern observed for $J/\psi$ at SPS and RHIC energies. In first type of models, one considers that $\chi_c$ and $\psi^{'}$ states evaporate shortly above $T_c$ while $J/\psi$ will do so if the temperature rises above $2.1\,T_c$ or $1.2\,T_c$ according to old~\cite{satz} and new~\cite{amocsy} estimates, respectively. Here one should emphasize that the new estimate for $T_D$ for $J/\psi$ poses an additional problem because then any enhancement in $J/\psi$ yield at RHIC energy would become difficult to explain. In second type of models, it is assumed that $J/\psi$ yield will result from a balance between annihilation of $J/\psi$ due to thermal gluons~\cite{xusat,gluon1} along with colour screening~\cite{mish,chu} and enhancement due to coalescence of uncorrelated $c\bar c$ pairs~\cite {grand,andro,thews} which are produced  at RHIC energy~\cite{adle,adar}. However, recent PHENIX data do not show a fully confirmed indication of $J/\psi$ enhancement except for the fact that $\langle p_T^2\rangle$ of the data and shape of rapidity-dependent nuclear modification factor $R_{AA}(y)$~\cite{pher} show some characteristics of coalescence production. 

    {\em Cold nuclear matter (CNM)} effects such as nuclear absorption, shadowing and anti-shadowing~\cite{vogt} are also expected to modify the $J/\psi$ yield. CNM effects due to the gluon shadowing and nuclear absorption of $J/\psi$ at the RHIC energy were evaluated from the $J/\psi$ measurement in d$+$Au collisions at RHIC~\cite{ads,adcnm}. Many pioneering workers have analyzed the SPS data~\cite{nac} on $J/\psi$ production (with regard to possible QGP effect) by employing ideas such as colour screening model~\cite{chu,karschf,ruus}, nuclear absorption model based on cut-off energy density~\cite{blaiz} etc. Chu and Matsui approach was originally designed to study the $p_T$ dependence by parameterizing the energy dependence in the transverse $z=0$ plane but they did not include the bag pressure term $B$ consistently. Karsch and Petronzio adopted essentially the same philosophy as Chu and Matsui without using bag constant and treating length of the plasma cylinder as a free parameter. Ruuskanen and Satz generalized this method also to include rapidity dependence by proposing a factorized formula $F(p_T,y)\approx f(p_T)\,g(y)$ for the fraction of deconfined $c\bar c$ pairs and studied its effect schematically. Blaizot and Ollitrault~\cite{blaiz} extended his nuclear absorption formulation~\cite{blaiz1} by assuming that no $J/\psi$ survives if produced in a region where the energy density (proportional to number of participants) exceeds a cut-off value. They focused only on the centrality dependence without considering hydrodynamical expansion of the plasma. The current PHENIX data on the centrality dependence of $J/\psi$ suppression have already been analyzed in several models such as comover model~\cite{cap}, statistical coalescence model~\cite{kost,gore}, kinetic model~\cite{grand}, statistical hadronization model~\cite{pbr} and QCD based nuclear absorption model~\cite{chow1}. None of these models gives a satisfactory description of the present experimental data. In particular, we do not find any single mechanism which can be used to explain the complete rapidity dependence of the $J/\psi$ suppression. Chaudhuri has attempted to explain the suppression at forward-rapidity~\cite{chow2} in a QGP motivated threshold model supplemented with normal nuclear absorption. However, this appears more like a parameter fitting because the parameters used in the analysis (e.g., QGP formation time $\tau=0.06-0.08$ fm/c) do not convey any physical meaning. Moreover, in order to accommodate rapidity dependence in the formulation, additional three parameters are used in the analysis e.g., $J/\psi$ nuclear absorption cross-section, threshold density and its smearing factor $\lambda$ to explicitly depend on the rapidity variable. Similarly Gunji et al.~\cite{gunji} have recently used a hydro$+J/\psi$ model in which QGP has been considered as undergoing (3+1)-dimensional hydrodynamic expansion but it has been used to explain the mid-rapidity data only without explicitly incorporating $J/\psi$ formation time. Recently, we analyzed~\cite{mish} the mid-rapidity PHENIX data of $J/\psi$ suppression, normalized by contribution due to cold nuclear matter effects~\cite{gunji,khar,raph}, using a modified colour screening model of Chu and Matsui~\cite{chu} by improving substantially the basic theme of refs.~\cite{chu,ruus}. More precisely, we had taken a bag model equation of state,  (1+1)-dimensional hydrodynamic evolution for QGP as well as additional time dilatation effect for charmonium as other ingredients in our model. We found that the model described the centrality dependence of $J/\psi$ suppression at mid-rapidity quite well~\cite{mish}. 

    In this paper, we generalize the above geometric formulation~\cite{mish} in order to incorporate the rapidity dependence by explicitly introducing rapidity variable in a consistent manner. Here again we consider (1+1)-dimensional hydrodynamic evolution of the QGP and the possible sequential dissociation scenario~\cite{khar} for the charmonium excited states. Our theory emphasizes the role of the medium's proper time in finding the locus of the screening region and also shows how this locus explicitly depends on the $J/\psi$ rapidity. We predict the centrality (i.e., impact parameter or number of participant nucleons $N_{part}$) dependence of the $J/\psi$ suppression in Au$+$Au collisions at mid-rapidity as well as at forward rapidity and compare with the recent experimental data of PHENIX collaboration at RHIC (normalized again with respect to CNM contribution). We notice that our model reproduces the main features of the data quite well. We also find that the present model gives more $J/\psi$ suppression at forward rapidity than that at mid-rapidity and this fact is again in quite good agreement with the recent PHENIX data. Results are reported corresponding to old~\cite{satz} and new~\cite{amocsy} set of values of the dissociation temperatures of the quarkonia. 
   

\section{Formulation}
Although our basic theme is similar to that described in refs.~\cite{mish,chu,dpal}, yet for the sake of convenience, we briefly recapitulate it below and also point out the important differences at appropriate places. For a QGP with massless quarks and gluons, the bag model EOS gives~\cite{mish,john}:
\begin{eqnarray}
  \epsilon = a\,T^4/c_s^2+B\quad;\quad P=a\,T^4-B\sim a\,(T^4-T_c^4);\\\nonumber
  c_s^2\equiv \frac{\partial P}{\partial\epsilon}\quad;\quad a\equiv \frac{37 \pi^2}{90}
  \quad;\quad B\equiv \frac{17\pi^2\,T_c^4}{45}\sim a\,T_c^4,
\end{eqnarray}
where at every time-space point $x\equiv(t,\vec x)$, $\epsilon$ is the energy density, $P$ the pressure, $T_c$ the critical temperature, $c_s^2$ the square of velocity of sound and $B$ is the bag constant. The coefficient $a$ is obtained by assuming an ideal gas of massless quarks and gluons with zero chemical potential and includes the number of degrees of freedom as a factor since the $\pi^2/90$ comes from integrating the thermodynamic distributions over momentum and is proportional to a value of Riemann-zeta function. It should be added here that the bag model EOS exhibits a first order phase transition between the QGP and the HG phases which is not well supported by the recent lattice QCD simulations~\cite{fod,kar}. Even though the bag model EOS has often been used for hydrodynamic calculations, it should be understood as a convenient way of parameterizing some features of the EOS with a rapid change of entropy density as a function of temperature in the transition region. For a QGP undergoing (1+1)-dimensional Bjorken's boost invariant expansion, the local thermodynamic observables become function of the lateral coordinate $r$ along with the proper time   $\tau\equiv(t^2-z^2)^{1/2}$, so that the cooling laws become
\begin{equation}
  \tilde{\epsilon}=\tilde{P}={\tilde{T}}^4={\tilde{\tau}}^{-q}\quad;\quad
  q\equiv 1+c_s^2,
\end{equation}
where the dimensionless symbols
  $\tilde{\tau}\equiv\tau/\tau_i$,\,$\tilde{T}\equiv
  T(\tau,r)/T(\tau_i,r)$,\, $\tilde\epsilon\equiv(\epsilon(\tau,r)-B)/(\epsilon(\tau_i,r)-B)$ and \, $\tilde{P}\equiv (P(\tau,r)+B)/(P(\tau_i,r)+B)$,
have been introduced for convenience with $\tau_i$ being the proper time for initial thermalization of the fireball.

    From (1) the pressure is seen to almost vanish at the transition point
$T_c$, i.e., in the hadronic sector. Hence on any transverse plane we choose the initial pressure profile
\begin{equation}
  P(\tau_i,r)=P(\tau_i,0)\,h(r)\quad;\quad h(r)\equiv \left(1-\frac{r^2}{R_T^2}\right)^{\beta}
  \theta(R_T-r),
\end{equation}
where $R_T$ denotes the radius of the cylinder. The power $\beta$ depends on the energy deposition mechanism, and $\theta$ is the unit step function. Clearly, our pressure is maximum at the center of the plasma but vanishes at the edge $R_T$ where hadronization occurs. The factor $P(\tau_i,0)$ is related to the mean pressure $<P>_i$ over the cross-section and to the corresponding average initial energy density $<\epsilon>_i$ via
\begin{equation}
  P(\tau_i,0)=(1+\beta)<P>_i=(1+\beta)\{c_s^2<\epsilon>_i-q\,B\}.
\end{equation}

   In the present work, we take the initial average energy density $<\epsilon>_i$ in terms of the number of participating nucleons $N_{part}$~\cite{khar} (which in turn depends on the impact parameter $b$), given by the modified Bjorken formula:
\begin{equation}
  <\epsilon>_i=\frac{\xi}{A_T\,\tau_i}\left(\frac{dE_T}{dy_H}\right)_{y_H=0}\quad;\quad A_T=\pi\,R_T^2,
\end{equation}
where $A_T$ is the transverse overlap area of the colliding nuclei, $(dE_T/dy_H)_{y_H=0}$ is the transverse energy deposited per unit rapidity of output hadrons. Both depend on the number of participants $N_{part}$~\cite{sscd} and thus provide centrality dependent initial average energy density $<\epsilon>_i$ in the transverse plane. The term $\xi$ is a phenomenological factor. The motivation and need of the $\xi$ factor will be discussed later in Sec. 3 in conjunction with the self-screened parton cascade model. 

   It is well known that $c\bar c$ bound state in a thermal medium feels a colour screened Yukawa potential and it melts at the dissociation  temperature $T_D$ (determined by quenched and full Lattice QCD simulations~\cite{satz,amocsy}) which corresponds to the energy density $\epsilon_s$ and pressure $P_s$ given by
\begin{equation}
  T_D\geq T_c\quad;\quad\epsilon_s=a\,T_D^4/c_s^2+B \quad;\quad P_s=a\,T_D^4-B.
\end{equation}
For any chosen fireball instant $t$ and on the arbitrary $z$ plane the contour of constant pressure
$P_s$ is obtained by combining the cooling laws (3) with the profile shape (4) to yield
\begin{equation}
 \tilde{P}\equiv\frac{P_s+B}{P(\tau_i,0)\,h(r)+B}={\tilde{\tau}}^{-q}.
\end{equation}
Setting $r=0$ the maximum  allowed tilde time $\tilde{\tau}_{s0}$ (during which pressure drops to $P_s$ at the center) can be identified as
\begin{equation}
  {\tilde{\tau}}_{s0}\equiv\left\{\frac{P(\tau_i,0)+B}{P_s+B}\right\}^{1/q},
\end{equation}
with $P(\tau_i,0)$ read-off from (4). Thereby the said locus takes the more convenient form
\begin{equation}
  \left(1-\frac{r^2}{R_T^2}\right)^{\beta}=H_s(\tau)\equiv\frac{{\tilde{\tau}}^q-B/(P_s+B)}{{\tilde{\tau}_{s0}}^q-B/(P_s+B)}.
\end{equation}

Our above result generalizes a similar expression derived by us~\cite{mish} for the special case when the transverse plane passes through the origin and the proper time was $t$ itself. 

      Next, consider an interacting $c\bar{c}$ pair created at the early time $t_1\sim \hbar/2m_c\,\approx 0$ at the location $(r_1,\phi_1,z_1)$ inside a cylinder of length $L_1$. The precise value of $L_1$ is not known a priori since different physical arguments can give different results, among which a recipe exploiting rapidity will turn out to be logical as discussed in the numerical section. The $c\bar c$ pair has mass $m$, transverse momentum $p_T$, transverse mass $m_T=\sqrt{m^2+p_T^2}$, rapidity $y$, total energy $p^0=m_T\,\cosh(y)$, longitudinal momentum $p_z=m_T\,\sinh(y)$, vectorial velocity $\vec v=(\vec p_T+\vec p_z)/p^0$ and dilation factor $\gamma=p^0/m$. In the fireball frame the pair will convert itself into the physical $J/\psi$ resonance after the lapse of time $t=\gamma\,\tau_F$ (with $\tau_F$ being the intrinsic formation time) provided the temperature $T<T_D$. At this instant the pair's transverse position $\vec r$, its longitudinal position $z$ and medium's proper time $\tau$ are given by
\begin{eqnarray}
\vec r=\vec r_1+v_T\,t\quad;\quad z=z_1+v_z\,t\\
\tau=(t^2-z^2)^{1/2}\,\theta(t-\mid z\mid)\quad;\quad t\equiv\gamma\tau_F.\nonumber
\end{eqnarray}
From the locus (9) we deduce the so called screening radius
\begin{equation}
  r_s=R_T\,\{1-H_s^{1/\beta}(\tau)\}^{1/2}\theta\{1-H_s(\tau)\}.
\end{equation}
In contrast to the earlier paper~\cite{mish}, the important role of proper time $\tau$ and hence of the longitudinal velocity $v_z$ must be noted in the definition of the screening radius $r_s$, which marks the boundary of the circular region where the quarkonium formation is prohibited. Since $\tau$ and hence $H_s(\tau)$ decreases as $|v_z|$ increases, it is clear that the radius $r_s$ of the screening region grows with growing rapidity. This fact will be utilized for interpreting our graphical results in Sec. 4. 

    Due to the existence of the screening region the pair will escape and form quarkonium if
 $\mid\vec{r_1}+\vec{v}_T\,t\mid\geq r_s$ implying 
\begin{equation}
  \cos(\phi_1)\geq C\quad;\quad C\equiv\frac
  {\left[(r_s^2-r_1^2)-v_T^2\,t^2\right]}{2\,r_1\,t\,|v_T|},
\end{equation}
where the role of the transverse velocity $v_T$ has become explicit. This trigonometric inequality is equivalent to saying that $-\phi_{max}(r_1,z_1)\leq\phi_1\leq\phi_{max}(r_1,z_1)$ where 

\[ \phi_{max}(r_1,z_1) = \left\{\begin{array}{clcr}
            \pi & \mbox{if $C \leq -1$}, \\
            \pi - \cos^{-1}\mid C\mid  & \mbox{if $-1\leq C \leq 0$}, \\
             \cos^{-1}\mid C\mid & \mbox{if $0\leq C \leq 1$}, \\
             0   & \mbox{if $C \geq 1$}.
\end{array}\right.\]
In contrast to our earlier work~\cite{mish} the symbol $\phi_{max}$ depends on both $r_1$ and $z_1$.

    Finally, we must deal with the anomalous survival probability due to QGP effect namely, $S(N_{part},p_T,y)$. Suppose the overall probability distribution for the production of $c\bar c$ pair at general position with general momentum is factorized as $P\propto f(r_1)\,g(p_T,y)$,       
where the radial profile function is
\begin{equation}
  f(r_1)\propto\left(1-\frac{r_1^2}{R_T^2}\right)^{\alpha}\theta(R_T-r_1),
\end{equation}
with $\alpha=0.5$~\cite{chu} and the momentum distribution $g(p_T,y)$ is left unspecified because it cancels out from the expression of the net survival probability in colour screening scenario defined by
\begin{equation}
S=\frac{\int_0^{R_T} dr_1\,r_1\,f(r_1)\int_{-L_1/2}^{L_1/2}dz_1\int_{-\phi_{max}}^{\phi_{max}}d\phi_1}
{\int_0^{R_T} dr_1\,r_1\,f(r_1)\,\int_{-L_1/2}^{L_1/2}dz_1\,\int_{-\pi}^{\pi}d\phi_1},
\end{equation}
which can be simplified as
\begin{equation}
S=\frac{2(\alpha+1)}{\pi\,R_T^2\,L_1}\int_0^{R_T} dr_1\,
r_1\,\left\{1-\left(\frac{r_1}{R_T}\right)^2\right\}^{\alpha}\int_{-L_1/2}^{L_1/2}\,dz_1\,\phi_{max}(r_1,z_1).
\end{equation}
In contrast to our previous paper~\cite{mish} above expression contains a non-trivial integration on $z_1$ coordinate. It must be emphasized that this generic symbol $S$ refers to a meson of given species and specified momentum.  

Often experimental measurement of $S$ at given $N_{part}$ or $y$ is reported in terms of the $p_T$ integrated yield ratio (nuclear modification factor) over the range $(p_T)_{min}\leq p_T\leq (p_T)_{max}$ whose theoretical expression would be
\begin{equation}
\langle S^{dir}\rangle=\frac{\int_{(p_T)_{min}}^{(p_T)_{max}}\,dp_T\,S^{dir}}{\int_{(p_T)_{min}}^{(p_T)_{max}}\,dp_T}.
\end{equation}
where the superscript "dir" pertains to the directly produced $J/\psi$ mesons.
 
    In nucleus-nucleus collisions~\cite{antoni}, it is known~\cite{khar} that only about $60\%$ of the observed $J/\psi$ originate directly in hard collisions while $30\%$ of them come from the decay of $\chi_c$ and $10\%$ from the $\psi^{'}$. Hence, the $p_T$ integrated "inclusive" survival probability $\langle S^{incl}\rangle$ of $J/\psi$ in the QGP becomes 
\begin{equation}
  \langle S^{incl}\rangle=0.6\,\langle S^{dir}\rangle_{\psi}+0.3\,\langle S^{dir}\rangle_{\chi_c}+0.1\,\langle S^{dir}\rangle_{\psi^{'}}.
\end{equation}
The hierarchy of {\it old} dissociation temperatures~\cite{satz} thus leads to sequential suppression pattern~\cite{karsf,gupta} with an early suppression of $\psi^{'}$ and $\chi_c$ decay products and much later one for the direct $J/\psi$ production. However, with {\it new} dissociation temperatures~\cite{amocsy} employing full lattice QCD (which almost mutually coincide) all the three species will show essentially the same suppression pattern i.e., the concept of sequential melting will not have any dramatic effect as will become evident later in Sec. 4.  

Now let us turn towards the numerical section of our work where both $T_D$ sets will be employed. 
\section{Numerical Work}
  Table 1 gives the values of various parameters used in our theory and the following explanations are relevant in this context. The value $T_c=0.17$ GeV is in accordance with the lattice QCD results~\cite{hats}. The choice $c_s^2=1/3$ is most common for free massless partons in an ideal gas, although for partons which carry thermal mass or interact among themselves, $c_s^2$ may be different like $1/5$~\cite{dpal}. The selection $\beta=1$~\cite{chu} indicates that the energy deposited in the collision is proportional to the number of nucleon-nucleon collisions. The initial proper time $\tau_i$ for QGP thermalization is taken as $0.6$ fm/c in accord with our earlier work~\cite{mish} as well as~\cite{hiran}. Also, relevant properties of the various quarkonia in a thermal medium are displayed in Table 2. The pressure $P_s$  corresponding to old and new  dissociation temperatures~\cite{satz,amocsy} for various quarkonium species are calculated from (6) and shown in Table 2. It is clear that, in going from $J/\psi$ to $\chi_c$ to $\psi^{'}$, the $T_D$ values (and hence also $P_s$) decrease some what sharply in the old set~\cite{satz} but rather slowly in the new set~\cite{amocsy}. As regards the length $L_1$ of the primordial cylinder (in $c=\hbar=1$ units) the Lorentz contracted total radius of the Au$+$Au system gives a small value $\sim 2\,R_{Au}/100\sim 0.14$ fm. The alternative estimate based on primordial creation time is also equally small namely, $\tau_1\sim 1/2m_c\sim 0.066$ fm/c which, however, can be made more logical by remembering that the corresponding longitudinal locations of the created $c\bar c$ pairs are expected to be of the order $|z_1|\sim \tau_1\,\sinh(|y|)$. Here $y$ is the rapidity whose experimentally determined bin reported in PHENIX experiment is $1.2\leq|y|\leq 2.2$. Hence the length $L_1$ of the cylinder is expected to lie in the range $0.10\leq L_1\leq 0.29$ so that its approximate mid value $L_1\sim 0.2$ fm can be used. 

Our numerical procedure proceeds through the following steps:\\
 (i) Before finding the centrality (or impact parameter) dependence of
$J/\psi$ suppression it is necessary to know the initial average energy density $<\epsilon>_i$ in terms of the number of participants $N_{part}$. For this purpose, we extract the transverse overlap area $A_T$ and the pseudo-rapidity distribution $(dE_T/d\eta_H)_{\eta_H=0}$ reported in ref.~\cite{sscd} at various values of  number of participants $N_{part}$. These $(dE_T/d\eta_H)_{\eta_H=0}$ numbers are then multiplied by a constant Jacobian 1.25 to yield the rapidity distribution $(dE_T/dy_H)_{y_H=0}$ occurring in (5).\\
(ii) As regards the choice of the $\xi$ parameter in (5) two routes are open. In the first route pertaining to the old $T_D$ values~\cite{satz} following point is noted. Even though the Bjorken formula provides a qualitatively good estimate of the initial energy density yet, it under-estimates the same, which can cause the suppression of only $\chi_c$ and $\psi^{'}$ but not of
$J/\psi$. Hence, a scaling-up factor $\xi=5$ has been introduced in (5) in order to obtain the desired $<\epsilon>_i=45$ GeV/fm$^3$~\cite{hiran} for most central collision. The relatively large values of our $<\epsilon>_i$ have the following justification : These are consistent with the predictions of the self-screened parton cascade model~\cite{eskola}, these also agree with the requirements of hydrodynamic simulation~\cite{hiran} which fit the pseudo-rapidity distribution of charged particle multiplicity $dN_{ch}/d\eta$ for various centralities already observed at RHIC, and these can cause melting of all the quarkonium species listed in Table 2. In the second route pertaining to the new set of $T_D$ values~\cite{amocsy} the original Bjorken's formula for $\langle\epsilon\rangle_i$ (without any $\xi$ parameter) is sufficient to cause the melting of all quarkonia species. The appropriate characterization of kinematic quantities in Au$+$Au collisions is presented in Table 3.\\ 
(iii) Next, we calculate the time $\tilde{\tau}_{s0}$ for the pressure to drop to $P_s$ at the
origin and thereby deduce the screening radius $r_s$ with the help of (8,9,10,11) for $J/\psi$ mesons of given $p_T$ and $y$.\\
(iv) Next, the quantity $C$ is computed from (12) which sets the condition for the quarkonium to escape from the screening region and the limiting values of the $\phi_{max}(r_1,z_1)$ are constructed using equation written just below (12).\\
(v) Finally, the survival probability $S$, at specified $y$ and fixed $p_T$ but varying $N_{part}$ is evaluated by Simpson quadrature using (15) from which $p_T$ integrated $\langle S^{dir}\rangle$ is also deduced.\\
(vi) The same numerical process is used to calculate the $\langle S^{dir}\rangle$ for all other higher quarkonia and finally, by including sequential melting of higher resonances, total $\langle S^{incl}\rangle$ has been calculated from (17).\\     
(vii) Since we have calculated the theoretical survival probability of $J/\psi$ due to the QGP effect, hence, for comparison with the actual experimental data at RHIC, following procedure is adopted. The experimental nuclear modification factor $R_{AA}$~\cite{pher} is divided by the contribution due to CNM effect $R_{AA}^{CNM}$~\cite{gunji,raph} so as to yield the experimental value of survival probability $S_{QGP}^{exp}$ due to QGP viz
\begin{equation}
  S_{QGP}^{exp}=\frac{R_{AA}}{R_{AA}^{CNM}},
\end{equation}
where $R_{AA}$ is the standard nuclear modification factor and $R_{AA}^{CNM}$ is a contribution to $R_{AA}$ originating from CNM effects constrained by the data of $d+Au$ collisions. It should be emphasized that deuteron-nucleus dA data are interesting both to fundamentally understand the issues of quarkonia as well as CNM and also to separate these effects from hot nuclear matter. Often the CNM effects are calculated by using theoretical model calculation~\cite{vogt} that include either EKS~\cite{eskola2} or NDSG~\cite{florian} shadowing models for the parton distribution functions : in each case an additional suppression associated with a $\sigma_{breakup}$ of $c\bar c$ pairs is also included. These methods of calculating $R_{AA}^{CNM}$ are thus model dependent. Uncertainties associated with these theoretical schemes can be avoided by using the so called data driven-method~\cite{raph}. In the data-driven method, the $R_{AA}^{CNM}$ is parameterized in terms of $R_{dA}(\pm y,b)$ as a function of impact parameter $b$ and mutually opposite rapidities $\pm y$ summed over the number of nucleon-nucleon collisions $N_{coll}$ calculated by Glauber model. These three methods are employed in ref.~\cite{adcnm} and also $R_{AA}$ in Au$+$Au as well as Cu$+$Cu collisions are compared with corresponding $R_{AA}^{CNM}$. 

    We take $R_{AA}^{CNM}$ from ref.~\cite{adcnm} calculated by data-driven method. Finally by using (18) our input data for the survival probability are generated. Of course, for a confirmed inference regarding this issue, error bars on the said ratio must be known. For convenience the mathematical expression for the error bar $\Delta_s$ in $S$ is recapitulated in the Appendix using individual standard deviation of the numerator and denominator. Unfortunately, the PHENIX experiments~\cite{pher,adcnm} do not tell what the correlation coefficient $r$ between $R_{AA}$ and $R_{AA}^{CNM}$ is. Assuming these quantities to be highly positively correlated i.e., $r=1$, we find
\begin{equation}
\Delta_s=S\Biggl\vert\frac{\Delta_{R_{AA}}}{R_{AA}}-\frac{\Delta_{R_{AA}^{CNM}}}{R_{AA}^{CNM}}\Biggr\vert.
\end{equation} 

\begin{figure}[tbp]
\begin{center}
\vbox{\includegraphics[scale=1.0]{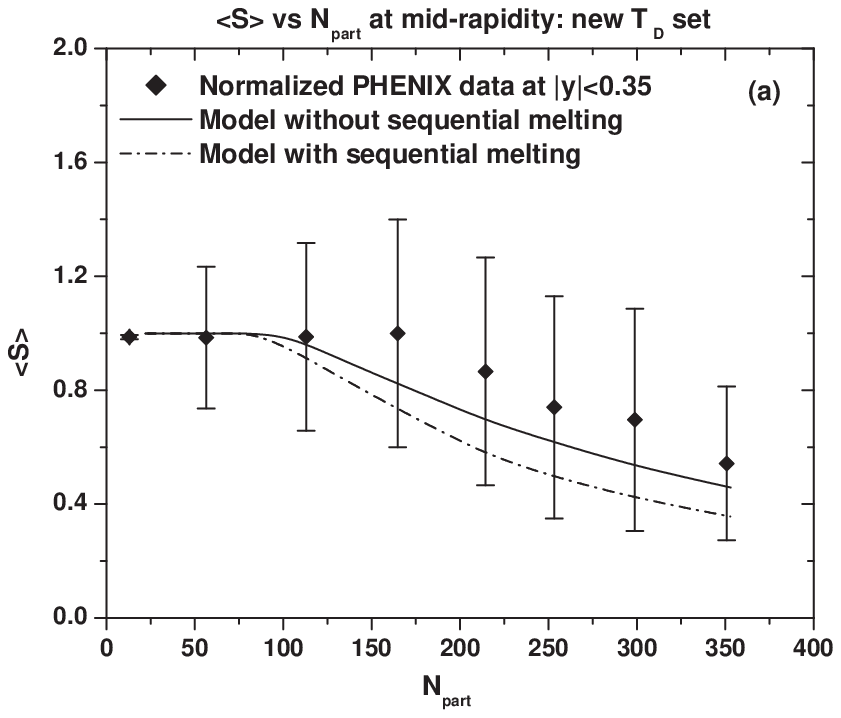}
\includegraphics[scale=1.0]{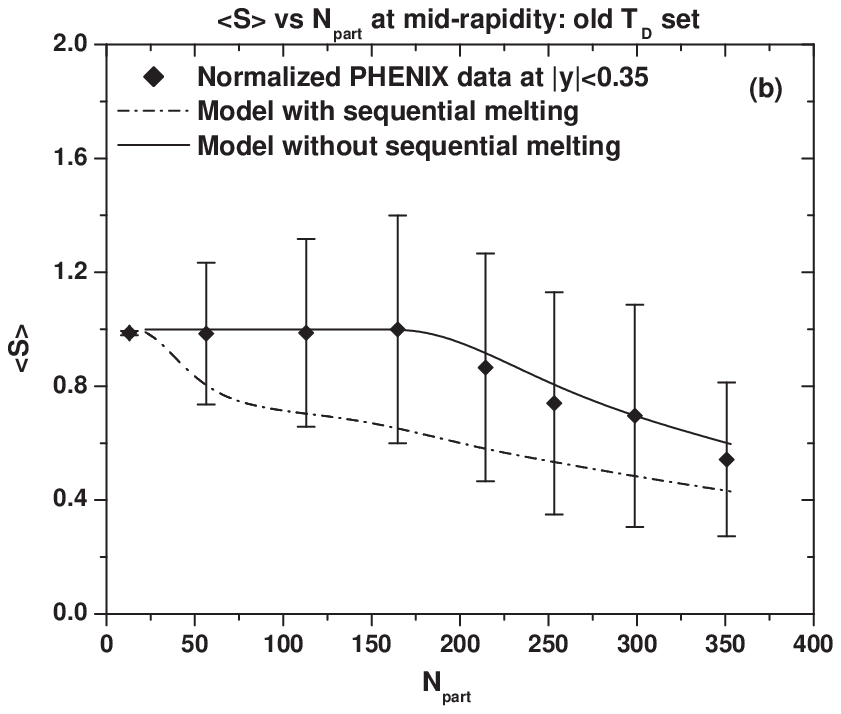}}
\caption{(a,b) The variation of $p_T$ integrated survival probability $\langle S\rangle$ (in the range allowed by invariant $p_T$ spectrum of $J/\psi$ measured by PHENIX experiment~\cite{pher,adcnm}) versus number of participants $N_{part}$ at mid-rapidity. The experimental data are shown by solid rhombus with error bars. In the top Figure, solid and dash dotted curves represent predictions of our present model without and with sequential melting using new set of $T_D$ values~\cite{amocsy}. In the bottom Figure, same are shown but by using old set of $T_D$ values~\cite{satz}. In both Figures $\tau_i=0.6$ fm/c is employed.}
\end{center}
\end{figure}

\begin{figure}[tbp]
\begin{center}
\vbox{\includegraphics[scale=1.0]{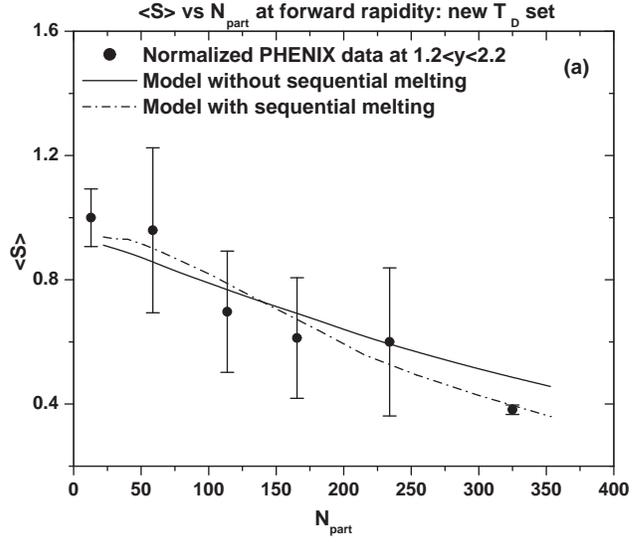}
\includegraphics[scale=1.0]{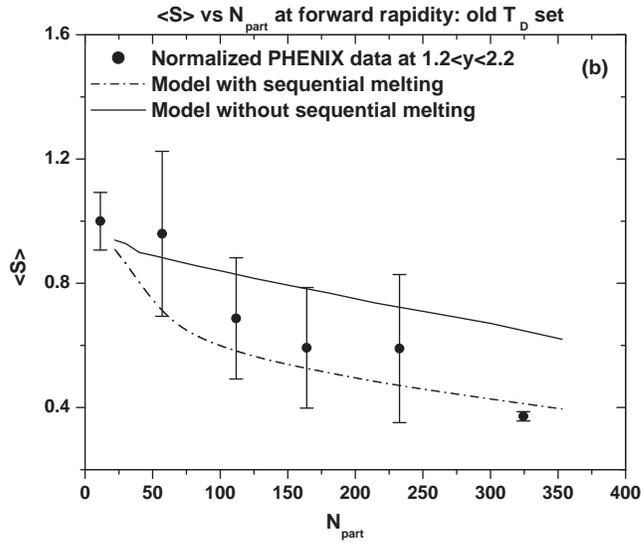}}
\caption{(a,b) Same as Fig. 1(a,b) but at forward rapidity~\cite{pher,adcnm}.}
\end{center}
\end{figure}

\begin{figure}[tbp]
\begin{center}
\vbox{\includegraphics[scale=1.0]{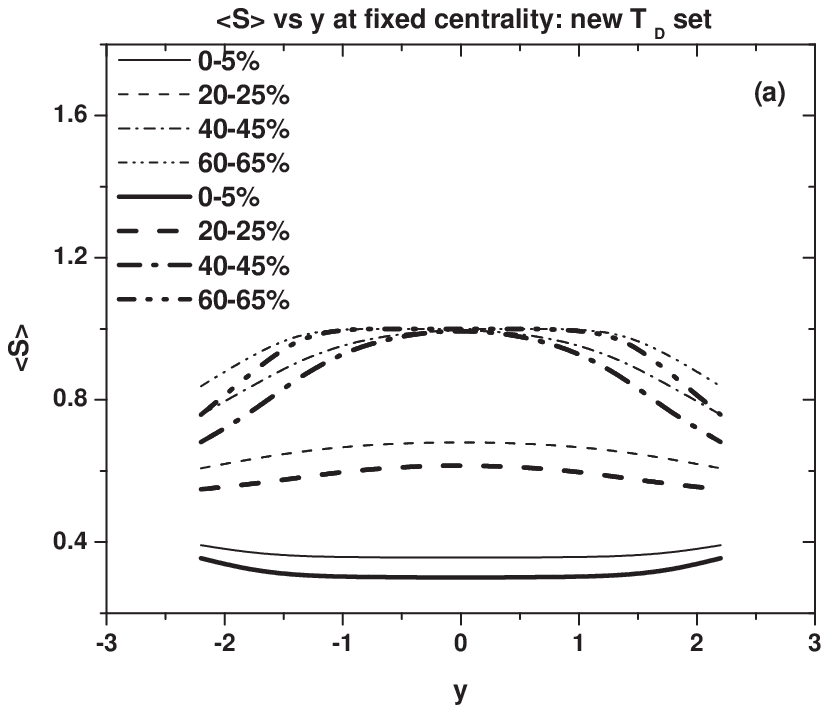}
\includegraphics[scale=1.0]{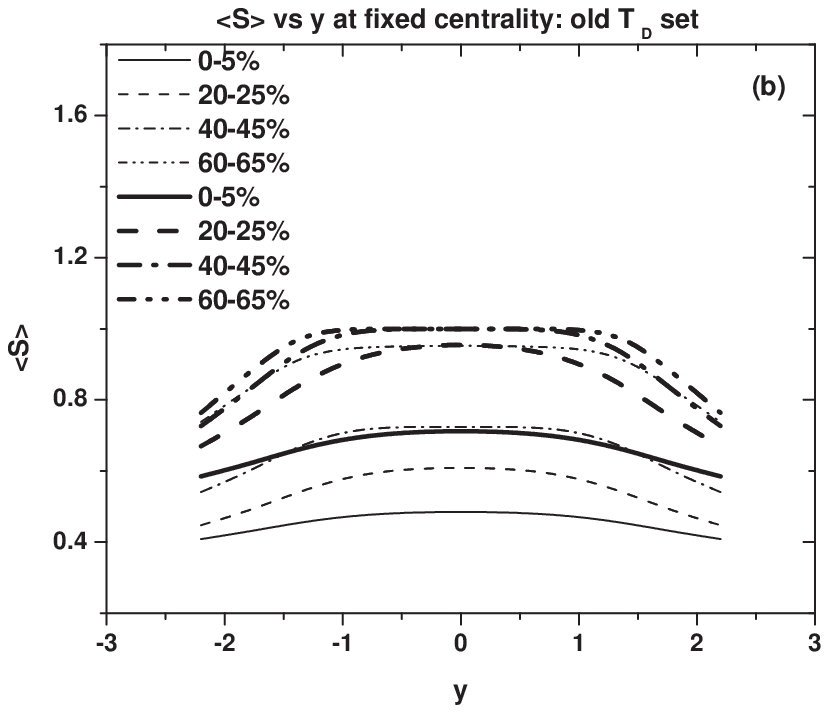}}
\caption{(a,b) Graph showing the theoretical variation of $p_T$ integrated survival probability i.e., $\langle S\rangle$ with respect to rapidity $y$ at fixed centralities. Top Figure corresponds to the results using new dissociation temperatures for quarkonia~\cite{amocsy} with and without sequential melting. Bottom Figure corresponds to the same as in top but by using old dissociation temperatures~\cite{satz}. In both Figures thin light curves and thick bold curves represent calculation with sequential melting and without sequential melting, respectively.}
\end{center}
\end{figure}


\begin{table}[tbp]
\caption{Various parameters used in the theory}
\begin{center}
\begin{tabular}{c|c|p{1.8cm}|c|c|c}
\hline\hline $T_c$ (GeV) & $c_s^2$ & $B$ (GeV/fm$^3$) & $\beta$ & $\alpha$ & $\tau_i$ (fm/c)\\
\hline 
$0.17$ & $1/3$ & $0.405$ & $1$ & $0.5$ & $0.60$\\
\hline\hline
\end{tabular}
\end{center}
\end{table}
\begin{table}[tbp]
\caption{Masses, formation times, dissociation temperatures~\cite{satz} and screening pressures $P_s$ of $J/\psi$, $\chi_c$ and $\psi^{'}$.}
\begin{center}
\begin{tabular}{l|l|l|l}
\hline\hline\mbox{} & $J/\psi$ & $\chi_c$ & $\psi^{'}$\\
 \hline $m$ (GeV) & 3.1 & 3.5 & 3.7\\ 
$\tau_F$ (fm)& 0.89 & 2.0 & 1.5 \\
 $T_D/T_c$ (Old) & 2.1 & 1.16 & 1.12 \\ 
 $T_D/T_c$ (New) & 1.2 & $\leq$ 1.0 & $\leq$ 1.0\\
 $P_s$ (Old) & 9.391  & 0.507 & 0.387\\
 $P_s$ (New) & 0.638  & 0.098  & 0.098\\
\hline\hline
\end{tabular}
\end{center}
\end{table}

\begin{table}[tbp]
\caption{Kinematic characterization of Au$+$Au collisions at RHIC at $\sqrt{s_{NN}}=200$ GeV~\cite{pher} with $\tau_i=0.6$ fm/c in Bjorken's formula when old (with $\xi=5$) and new sets of $T_D$ (without $\xi$ factor) values~\cite{satz,amocsy} are used.}
\begin{center}
\begin{tabular}{c|c|p{4.0cm}|p{4.2cm}}
\hline\hline 

$N_{part}$ & $R_T$ (fm) & $<\epsilon>_i$ (GeV/fm$^3$) (With $\xi$ factor in old $T_D$ sets) & $<\epsilon>_i$ (GeV/fm$^3$) (Without $\xi$ factor in new $T_D$ sets) \\\hline
      22.0   & 3.45 & 5.86  &  1.17\\
      30.2   & 3.61 & 7.92  &  1.58\\
      40.2   & 3.79 & 10.14 &  2.03\\
      52.5   & 3.96 & 12.76 &  2.55\\
      66.7   & 4.16 & 15.69 &  3.14\\
      83.3   & 4.37 & 18.58 &  3.72\\
      103.0  & 4.61 & 21.36 &  4.27\\
      125.0  & 4.85 & 24.38 &  4.88\\
      151.0  & 5.12 & 27.37 &  5.47\\
      181.0  & 5.38 & 30.52 &  6.10\\
      215.0  & 5.64 & 34.17 &  6.83\\
      254.0  & 5.97 & 37.39 &  7.48\\
      300.0  & 6.31 & 41.08 &  8.22\\
      353.0  & 6.68 & 45.09 &  9.02\\
\hline\hline
\end{tabular}
\end{center}
\end{table}
Now we turn to physical interpretations of our results.
\section{Results and Discussions}

   In this section we shall present/discuss our numerical results remembering the fact that the cooling laws in (2) and hence the radius of screening regions in (11) depend on the proper time $\tau$. In turn when the $J/\psi$ meson's trajectory is also considered in accordance with (10) we have to distinguish between the cases of mid-rapidity ($v_z\approx 0$) and forward rapidity ($v_z > 0$) separately.

{\nd \it Scenario at mid-rapidity}
  
  Figure 1(a,b) show the variation of $p_T$ integrated survival probability $\langle S\rangle$ with respect to number of participants $N_{part}$ at mid-rapidity corresponding to the new and old $T_D$ sets, respectively. The three curves on each Figures correspond to experimental data $S_{QGP}^{exp}$ (cf.18; solid rhombus), our model calculation without sequential melting $\langle S_{J/\psi}^{dir}\rangle$ (cf.16; solid line) and that with sequential melting $\langle S^{incl}\rangle$ (cf.17; dash dotted line), respectively. It is obvious from both Figures that $\langle S\rangle$ decreases with increase in $N_{part}$ both experimentally and theoretically because of the increase in energy density with respect to $N_{part}$ inside the screening region. In Figure 1(a) both $\langle S_{J/\psi}^{dir}\rangle$ and $\langle S^{incl}\rangle$ are in qualitatively good agreement with the experimental data although quantitatively $\langle S^{incl}\rangle$ is slightly underestimated. The tiny difference between $S_{J/\psi}^{dir}$ and $S^{incl}$ shows that the sequential melting mechanism does not have any dramatic influence here due to a very small difference in the dissociation temperatures of the quarkonia species in the new estimate for $T_D$. Let us now pay attention to Figure 1(b) based on the old $T_D$ set where a parameter $\xi=5$ was needed to scale up the initial energy density. The agreement between $S_{J/\psi}^{dir}$ and $S_{QGP}^{exp}$ is excellent while that between $S^{incl}$ and $S_{QGP}^{exp}$ is definitely poor. The reason behind this observation is that since the dissociation temperatures of $J/\psi$ is far above that of $\chi_c$ and $\psi^{'}$, the latter two species just break-up before they can possibly decay to produce secondary $J/\psi$'s. In other words, the directly produced $J/\psi$'s are responsible for the observed survival pattern with sequential melting not playing any significant role. Furthermore, the difference between our theoretical predictions with and without sequential melting is substantial due to large difference between the dissociation temperatures of quarkonium species in the calculations with an old set of $T_D$ values. 
    
{\nd \it Scenario at forward rapidity}

Figures 2(a,b) are also similar to Figures 1(a,b) except that they pertain to forward rapidity. The main difference between Figures 1(a,b) and Figures 2(a, b) arises from the fact that at fixed $N_{part}$ but non-zero rapidity ($v_z > 0$), $\langle S\rangle$ decreases with increasing rapidity. This is understandable because, as pointed out beneath (11) the radius $r_s$ of the screening region at higher rapidity is larger than that at lower rapidities. Therefore, for a $J/\psi$ meson of given $p_T$, the traversal time will also be correspondingly larger implying that the predicted survival probability must decrease with increasing $|y|$. Also, in Figure 2(a) both the curves of $\langle S_{J/\psi}^{dir}\rangle$ and $\langle S^{incl}\rangle$ agree very well with $S_{QGP}^{exp}$ and reasonably well in Figure 2(b). Hence based on the analysis at forward rapidity, it is not possible to make a preferential choice between the $\langle S_{J/\psi}^{dir}\rangle$ and $\langle S^{incl}\rangle$ models.  

{\nd \it Detailed rapidity dependence}
 
  Figures 3(a,b) show $\langle S^{dir}\rangle$ and $\langle S^{incl}\rangle$ plotted versus rapidity $y$ at fixed centralities. The chosen centralities are 0-5\%, 20-25\%, 40-45\% and 60-65\%. These centralities do not exactly correspond to the recent PHENIX measurement~\cite{pher} on $J/\psi$ suppression. Rather they correspond to an earlier measurement~\cite{sscd} in which the relevant $(dE_T/d\eta_H)_{\eta_H=0}$ values versus $N_{part}$ were given for insertion in Bjorken formula. It is clear from Figures 3(a,b) that $\langle S\rangle$ shows flat maximum around $|y|\approx 0$ and decreases for large $y$ as explained above. This trend qualitatively agrees with the experimental data~\cite{pher} but for doing quantitative comparison with the data, calculations will have to be done at exactly same centralities reported in ref.~\cite{pher}. Furthermore, the difference between the predicted values of $\langle S\rangle$ (both $\langle S^{dir}\rangle$ and $\langle S^{incl}\rangle$) between mid and forward rapidities diminishes in going from non-central to central collisions. This trend again agrees with the recent PHENIX data~\cite{pher}. A satisfying feature of Figures 3(a,b) is that our model predictions for $\langle S\rangle$ are obtained by simply inserting the relevant values of the rapidity $y$ in contrast to other model~\cite{chow2} where different additional parameters are needed for describing the data in different rapidity bins. 
  
It should be noted that our model does not involve any free parameter although there are some parameters used in the calculation. We have assigned proper justifications to their values and these have also been used in other calculations e.g., in Chu and Matsui model~\cite{chu} and in hydrodynamical models~\cite{gunji,hiran}. 

At this stage a comment is warranted on the uncertainties involved in the CNM effects~\cite{adcnm} extracted by data-driven method and constrained by d$+$Au collision measurements. In view of the relatively large error bars in $R_{AA}^{CNM}$ reported in ref.~\cite{adcnm}, it is not possible to make any firm quantitative statements on any additional $J/\psi$ suppression in Au$+$Au collisions beyond that expected from cold nuclear matter effects (i.e., due to possible QGP formation). We feel, therefore, that more firm conclusion about possible QGP formation can be drawn by analyzing results such as Figures(1-3) if more accurate measurement of $R_{dA}$ is done in future experiments. 
 
\section{Summary and Conclusions}

     Generalization of our previous work~\cite{mish}, based on the $J/\psi+$hydro framework involving (1+1)-dimensional expansion of Chu and Matsui~\cite{chu} model, has been done in the present paper in order to incorporate complete rapidity dependence of the $J/\psi$ suppression. The present work also takes into account the additional time dilatation effect for charmonium formation due to its motion along a general direction and this makes our model different from other calculations e.g., by Chaudhari~\cite{chow2} and by Gunji et al.~\cite{gunji}. Our formulation for $J/\psi$ suppression shows explicit dependence on transverse momentum, centrality as well as on the rapidity and the predictions are consistent with the data. Our model has suitably incorporated the results from a recent lattice simulation on the dissociation temperatures~\cite{satz,amocsy} and formation time of $J/\psi$, as well as on the non-sequential/sequential melting of higher resonances. As expected, using new $T_D$ values we find that within error bars, the predictions without sequential decays of higher charmonia states provide better results. We conclude from our results that sequential decay mechanisms usually invoked for explaining the similarity of $J/\psi$ suppression pattern at SPS and RHIC, is simply ruled out by our present model. The conclusion remains surprisingly valid even in the case of the old values of $T_D$ used in the calculation for different charmonia states. It appears that the presence of a large initial energy density in this case makes the survival of various charmonia states difficult one and consequently the feed down factor from higher charmonia states does not improve the situation. Thus our calculation over emphasizes the role of production and dissociation of direct $J/\psi$ in the QGP scenario. We have analyzed the centrality dependence of the $J/\psi$ suppression data available from RHIC (normalized by contribution due to CNM effects) in terms of the survival probability versus number of participants at mid-rapidity as well as at forward rapidity. Our results reproduce the main features of the PHENIX data since we observe more suppression at forward rapidity as compared to mid-rapidity region and this is consistent with the recent PHENIX data. It is creditable to notice that the same mechanism explains the suppression patterns observed in the entire rapidity regions. Further extension of this work, such as incorporation of (3+1)-dimensional hydrodynamic expansion and the predictions of our model at LHC energies would provide additional support to our ideas.

  In conclusion, $J/\psi$ suppression can still be regarded as one of the most potential signatures for deconfinement. Recent PHENIX data on $J/\psi$ suppression and theoretical Lattice estimates for the dissociation temperatures of $J/\psi$ family have added more confusions regarding the interpretation of the data. Does a low depletion of $J/\psi$ yields at RHIC support a much stronger direct $J/\psi$ suppression (at dissociation temperatures close to $T_c$) and some other mechanism like abundant charm quark production thus creating more $c\bar c$ pairs ? More precise data for d$+$Au collisions are needed to draw a firm conclusion on this issue. We believe that the LHC measurements of $J/\psi$ yield in Pb$+$Pb collisions at $5.5$ TeV will help in resolving some of these confusing issues like sequential screening scenario and dissociation behaviours of $J/\psi$ and/or $\Upsilon$ families. 

\begin{acknowledgments}
M. Mishra and V. J. Menon are grateful to the Council of Scientific and Industrial Research~(CSIR), New Delhi for their financial assistance. C. P. Singh acknowledges the financial support from a research project sanctioned by Department of Science and Technology (DST), New Delhi. We gratefully acknowledge the helpful comments from Dr. Rapha\"{e}l Granier de Cassagnac on this work. In fact he had suggested us the problem in a private communication.
\end{acknowledgments}

\section*{APPENDIX : ERROR BAR CALCULATION IN S (cf.(18))}
  Let {\em a} be a random variate with mean $\langle a\rangle$, fluctuation $\delta_a=a-\langle a\rangle$, and standard deviation $\Delta_a=\langle \delta_a^2\rangle^{1/2}$. Similar quantities for another variate $b$ are called $\langle b\rangle$, $\delta_b$ and $\Delta_b$, respectively. The covariance between $a$ and $b$ is denoted by $C_{ab}=\langle\delta_a\,\delta_b\rangle=r\Delta_a\Delta_b$ with the correlation coefficient $r$ lying in the range $-1\leq r\leq +1$. The ratio $S$ between the variates has the statistical properties
$$
\begin{array}{lcl}
S=\frac{a}{b}\quad;\quad \frac{\delta_s}{S}=\frac{\delta_a}{a}-\frac{\delta_b}{b}
\end{array}\eqno(A1)
$$  
$$
\begin{array}{lcl}
\biggl\vert\frac{\Delta_S}{S}\biggr\vert=\langle(\frac{\delta_S}{S})^2\rangle^{1/2}=\left\{\left(\frac{\Delta_a}{a}\right)^2+\left(\frac{\Delta_b}{b}\right)^2-2\,\frac{\Delta_a}{a}\frac{\Delta_b}{b}\,r\right\}^{1/2}
\end{array}\eqno(A2)
$$
whose special values read
$$
\begin{array}{lcl}
\biggl\vert\frac{\Delta_S}{S}\biggr\vert=\Biggl\vert\biggl\vert\frac{\Delta_a}{a}\biggr\vert-\biggl\vert\frac{\Delta_b}{b}\biggr\vert\Biggr\vert\quad;\quad\biggl\vert\frac{\Delta_a}{a}\biggr\vert+\biggl\vert\frac{\Delta_b}{b}\biggr\vert\quad;\quad 
\left\{\left(\frac{\Delta_a}{a}\right)^2+\left(\frac{\Delta_b}{b}\right)^2\right\}^{1/2}
\end{array}\eqno(A3)
$$
according as $r=+1,\,-1,\,0$, respectively.

In the problem of $J/\psi$ suppression the error bar formula (19) is obtained by identifying $a=R_{AA}$, $b=R_{AA}^{CNM}$ and assuming $r=1$.

\end{document}